\documentclass[aps,pra,twocolumn,nofootinbib,preprintnumbers,english]{revtex4-1}
\usepackage{amsthm,amsmath,amsfonts,dsfont,upgreek}    
\usepackage{amssymb,epsfig,setspace}
\usepackage{graphicx}
\usepackage{babel}
\usepackage{verbatim}

\begin{document}


\hyphenation{ano-ther ge-ne-ra-te dif-fe-rent know-le-d-ge po-ly-no-mi-al}
\hyphenation{me-di-um  or-tho-go-nal as-su-ming pri-mi-ti-ve pe-ri-o-di-ci-ty}
\hyphenation{mul-ti-p-le-sca-t-te-ri-ng i-te-ra-ti-ng e-q-ua-ti-on}
\hyphenation{wa-ves di-men-si-o-nal ge-ne-ral the-o-ry sca-t-te-ri-ng}
\hyphenation{di-f-fe-r-ent tra-je-c-to-ries e-le-c-tro-ma-g-ne-tic pho-to-nic}
\hyphenation{Ray-le-i-gh di-n-ger Kra-jew-ska Wal-czak Ham-bur-ger Ad-di-ti-o-nal-ly}
\hyphenation{Kon-ver-genz-the-o-rie ori-gi-nal in-vi-si-b-le cha-rac-te-ri-zed}
\hyphenation{Ne-ver-the-less sa-tu-ra-te Ene-r-gy sa-ti-s-fy prac-ti-cal im-ple-men-ta-tion}
\hyphenation{ap-pear down-loa-ded nu-me-rics va-ni-shes mi-ni-mal}

\title{A hidden analytic structure of the Rabi model}

\author{
Alexander Moroz} 

\affiliation{Wave-scattering.com}  
 
\begin{abstract}
The Rabi model describes the simplest interaction between a 
cavity mode with a frequency $\omega_c$ and a two-level system 
with a resonance frequency $\omega_0$. It is shown here 
that the spectrum of the Rabi model coincides with the support 
of the discrete Stieltjes integral measure in the orthogonality 
relations of recently introduced orthogonal polynomials.
The exactly solvable limit of the Rabi
model corresponding to $\Delta=\omega_0/(2\omega_c)=0$, which describes a displaced harmonic 
oscillator, is characterized by the discrete Charlier polynomials in normalized energy
$\upepsilon$, which are orthogonal on an equidistant lattice. 
A non-zero value of $\Delta$ leads to non-classical 
discrete orthogonal polynomials $\phi_{k}(\upepsilon)$ and induces a deformation
of the underlying equidistant lattice. The results provide a basis 
for a novel analytic method of solving the Rabi model. The number 
of ca. {\em 1350} calculable energy levels per parity subspace 
obtained in double precision (cca 16 digits) by an elementary stepping algorithm 
is up to two orders of magnitude higher than is possible to 
obtain by Braak's solution. Any first $n$ eigenvalues of the 
Rabi model arranged in increasing order can be determined as zeros 
of $\phi_{N}(\upepsilon)$ of at least the degree $N=n+n_t$. 
The value of $n_t>0$, which is slowly increasing with $n$, 
depends on the required precision. For instance, $n_t\simeq 26$ 
for $n=1000$ and dimensionless interaction constant $\kappa=0.2$, 
if double precision is required. 
Given that the sequence of the $l$th zeros $x_{nl}$'s 
of $\phi_{n}(\upepsilon)$'s defines a monotonically
decreasing discrete flow with increasing $n$, the Rabi model is 
indistinguishable from an algebraically solvable model in any 
finite precision. Although we can rigorously prove our results 
only for dimensionless interaction constant $\kappa< 1$, 
numerics and exactly solvable example suggest 
that the main conclusions remain to be valid also for $\kappa\ge 1$.
\end{abstract}

\pacs{03.65.Ge, 02.30.Ik, 42.50.Pq}

\maketitle 

\section{Introduction}
\label{sc:intr}
Let us consider a quantum model described by
a Hamiltonian $\hat{H}$ satisfying the eigenvalue
equation 
\begin{equation}
\hat{H}\upvarphi=E\upvarphi
\label{eme}
\end{equation}
in the Bargmann Hilbert space $\mathfrak{b}$
of analytic entire functions \cite{Schw,Brg}.
The latter implies that any physical state is 
described by an entire function 
\begin{equation}
\upvarphi(z)=\sum_{n=0}^\infty \phi_n z^n,
\label{pss}
\end{equation}
where $\{\phi_n\}_{n=0}^\infty$ are 
the sought expansion coefficients. 
The present work investigates consequences of the following
three simple observations.

{\em First observation}.--
A first trivial observation is that for $\upvarphi$ to be an 
element of $\mathfrak{b}$, the coefficients $\phi_n$'s
have to approach zero in the limit $n\rightarrow\infty$. 
Hence for energy $\upepsilon$ to belong to the spectrum
$\Upsigma$, $\phi_n$ have to be such a solution of
Eq. (\ref{eme}) that $\phi_n\rightarrow 0$. Briefly,
\begin{equation}
\upepsilon\in \Upsigma \Longrightarrow \phi_n\rightarrow 0~~~~~
(n\rightarrow\infty).
\label{cd1}
\end{equation}
Obviously, the arrow {\em cannot} be reversed without
some further limitations. There could be solutions 
of Eq. (\ref{eme}) going to zero in the limit $n\rightarrow\infty$ 
which need not lead to an entire function, and 
hence to an element of $\mathfrak{b}$ 
(e.g. $|\phi_n|\sim n^{-c}$, where $c$ 
is an arbitrary positive constant).

{\em Second observation}.--
For a number of models \cite{Schw,AMep,Zh1}, the eigenvalue
equation (\ref{eme}) reduces in the Bargmann space 
$\mathfrak{b}$ to a {\em three-term difference equation}
\begin{equation}
\phi_{n+1} + a_n \phi_n + b_n \phi_{n-1}=0 \hspace*{1.8cm} (n\ge 0).
\label{3trg}
\end{equation}
The recurrence coefficients $a_n$ and $b_n$ are functions 
of model parameters, and so are the coefficients $\phi_n$'s.
Our {\em second} observation regards the case when
(i) $b_n\neq 0$ and 
(ii) the recurrence coefficients
have at most an asymptotic power-like dependence 
\begin{equation}
a_n\sim a n^{\varsigma},~~~~~~~ b_n\sim b n^{\upsilon}
~~~~~~~~~~~~~~ (n\rightarrow\infty),
\label{rcd}
\end{equation}
where $2\varsigma>\upsilon$ and $\tau=\varsigma-\upsilon\geq 1/2$. 
The above conditions select an important class ${\cal R}$
of quantum models that was initially introduced and studied 
in our earlier work \cite{AMep}. Prominent examples 
comprise a displaced harmonic oscillator \cite{Schw,AMep,AMops}, the
Rabi model \cite{Rb}, two-mode squeezed 
harmonic oscillator \cite{Zh1}, etc. 
The {\em second} observation is that, for the models of 
${\cal R}$, also the reverse condition to that in
Eq. (\ref{cd1}) applies. 
Energy $\upepsilon$ belongs to the spectrum
$\Upsigma$ if and only if $\phi_n(\upepsilon)\rightarrow 0$ in the limit
$n\rightarrow\infty$. We have the spectral condition
\begin{equation}
\upepsilon\in \Upsigma \Longleftrightarrow 
          \phi_n(\upepsilon)\rightarrow 0 \hspace*{1.2cm}
(n\rightarrow\infty).
\label{cd2}
\end{equation}
Indeed, according to
the Perron-Kreuser theorem (Theorem 2.3 in Ref. \cite{Gt}),
there are possible two qualitatively 
different types of linearly independent 
solutions of the recurrence (\ref{3trg}). 
The asymptotic behaviour of 
the {\em minimal} solution 
guaranteed by the Perron-Kreuser theorem is
\begin{equation}
\frac{\phi_{n+1}}{\phi_n}\sim 
     -\frac{b}{a}\frac{1}{n^\tau} \rightarrow 0
\hspace*{1.2cm} (n\rightarrow \infty)
\label{mins}
\end{equation}
[in virtue of Eq. (\ref{rcd}) and $\tau \geq 1/2>0$].
On the other hand, the {\em dominant} 
solutions of the recurrence (\ref{3trg}) behave 
as $\phi_{n+1}/\phi_n \sim -an^\varsigma$
in the limit $n\rightarrow \infty$. For either (i) $\varsigma>0$
or (ii) $\varsigma=0$ and $a> 1$ the absolute value of 
$\phi_n$ tends to infinity. The above dichotomy precludes 
any intermediate behaviour like 
$|\phi_n|\sim n^{-c}$.
Consequently, any solution of the 
recurrence (\ref{3trg}) with
given initial conditions 
that behaves as $\phi_n(\upepsilon)\rightarrow 0$ 
in the limit $n\rightarrow \infty$
corresponds necessarily to an eigenvalue $\upepsilon\in\Upsigma$.

{\em Third observation}.--
Our {\em third} observation concerns the case 
when each of the expansion coefficients $\phi_n$'s 
is proportional to a polynomial of degree $n$ in the energy 
parameter $\upepsilon$ \cite{AMops}. 
We recall that the {\em necessary} and {\em sufficient} 
condition for a family of polynomials $\{P_n\}$ (with
degree $P_n = n$) to form an {\em orthogonal polynomial system} (OPS)
is that $P_n$'s satisfy 
\begin{equation}
P_n(x)=(\beta_n x -c_n)P_{n-1}(x) - \lambda_n P_{n-2}(x)
\label{chi3trg}
\end{equation}
with the initial condition $P_{-1}(x)=0$ and $P_{0}(x)=1$,
where the coefficients $\beta_n,\, c_n$ 
and $\lambda_n$ are independent
of $x$, $\beta_n\ne 0$, and 
$\lambda_n\ne 0$ for $n\ge 1$ \cite{Chi}.
In what follows, the Stieltjes measure $d\psi(x)$ 
in the orthogonality relations induced by the positive moment functional 
\begin{eqnarray}
 {\cal L}[P_m(x)P_n(x)] & = & 
\int_{-\infty}^\infty P_m(x)P_n(x)\, d\psi(x) 
\nonumber\\
& = & \lambda_1\lambda_2\ldots
\lambda_{n+1}\delta_{mn},
\label{bor}
\end{eqnarray}
shall have a {\em discrete} support \cite{AMops}. 
In other words the set of all points $x$ at which the 
resulting Stieltjes measure $d\psi(x) \equiv \psi(x) - \psi(x-0)>0$ 
forms a discrete lattice $\Lambda$ \cite{Chi}.
(In mathematics, the set $\Lambda$ is called the spectrum of $\psi$
- cf. p. 51 of Ref. \cite{Chi}.)

Orthogonal polynomials that are pairwise orthogonal 
with respect to a discrete measure are called {\em discrete} 
orthogonal polynomials \cite{NSU,KS}. The requirement 
that the expansion coefficients $\phi_n$'s are proportional to
{\em discrete} orthogonal polynomials obviously entails a certain restriction
on the form of the recurrence coefficients $a_n$ and $b_n$ of our
initial recurrence (\ref{3trg}). 
Nevertheless, the latter is still satisfied 
for a number of important models \cite{AMops}.
Canonical properties of an OPS are that $P_n$'s

\begin{itemize}

\item have {\em real} and {\em simple} 
zeros (Theorem I-5.2 of Ref. \cite{Chi}), 

\item the zeros of any two subsequent 
polynomials $P_n(x)$ and $P_{n+1}(x)$ 
mutually separate each other 
(Theorem I-5.3 of Ref. \cite{Chi}).

\end{itemize}
Specifically, denote the zeros of $P_n(x)$ with degree 
$P_n = n$ by $x_{n1}<x_{n2}<\ldots< x_{nn}$.
Then for any $l=1,2,\ldots,\allowbreak n-1$
\begin{equation}
x_{nl}<x_{n-1,l}<x_{n,l+1}.
\label{1p5p4}
\end{equation}
For each fixed $l$, $\{x_{nl}\}_{n=l}^\infty$ is a strictly 
{\em decreasing} sequence, defining a discrete flow, and the limit 
\begin{equation}
\lim_{n\rightarrow\infty} x_{nl} = \xi_l \in \Upsigma 
\label{1p5p6}
\end{equation}
exists \cite{Chi}. Because of the spectral condition (\ref{cd2}), 
the discrete flow has nowhere to flow than 
to the spectral point of our model. 
Hence $\xi_l\in\Upsigma$. 
In other words, the spectrum of a
physical model coincides with the corresponding discrete lattice $\Lambda$.

{\em Central questions}.--
The first two observations might have prompted the 
knowledgeable reader to pose the following question {\bf Q1}: 
{\em Could the zeros of a given
$\phi_n(\upepsilon)$
be of use to determine the spectrum of models 
from the recurrence class ${\cal R}$}?

The very fact that the spectrum of a model is
determined as zeros of a polynomial implies
a special case of analytic solvability known as
{\em algebraic} solvability \cite{TU,Trb,BD,KUW,KKT}.
Our another question is therefore {\bf Q2}:
{\em Are there some models of the class ${\cal R}$ 
which are algebraically solvable}? 

The rest of the paper
is devoted to answering the questions in the special
case of the Rabi model \cite{Rb}.

\section{Rabi model}
\label{sc:rbm}
The Rabi model \cite{Rb} describes the 
simplest interaction between a 
cavity mode with a frequency $\omega_c$ and a two-level system 
with a resonance frequency $\omega_0$. 
The model is characterized by the Hamiltonian \cite{Schw,Rb} 
\begin{equation}
\hat{H}_R =
\hbar \omega_c \mathds{1} \hat{a}^\dagger \hat{a} 
 + \hbar g\sigma_1 (\hat{a}^\dagger + \hat{a}) + \mu \sigma_3,
\label{rabih}
\end{equation}
where $\mu=\hbar \omega_0/2$, $\hat{a}$ and $\hat{a}^\dagger$ are 
the conventional boson annihilation and creation operators 
satisfying commutation relation 
$[\hat{a},\hat{a}^{\dagger}] = 1$, and $g$ is a coupling constant. 
In what follows, $\mathds{1}$ is the unit matrix,
$\sigma_j$ are the Pauli matrices in 
their standard representation, and we set the 
reduced Planck constant $\hbar=1$.
The Hilbert space is 
${\cal B}=L^2(\mathbb{R})\otimes\mathbb{C}^2$, where
$L^2(\mathbb{R})$ is represented by the Bargmann space of 
entire functions $\mathfrak{b}$, and $\mathbb{C}^2$ 
stands for a spin space \cite{Schw,Brg}.
In a unitary equivalent {\em single-mode spin-boson picture}, 
$\hat{H}_R$ becomes
\begin{equation}
\hat{H}_{sb} =
\omega_c \mathds{1} \hat{a}^\dagger \hat{a} + \mu \sigma_1 
        +  g \sigma_3 (\hat{a}^\dagger + \hat{a}).
\end{equation}
The transformation is accomplished 
by means of the unitary operator 
\begin{equation}
U= \frac{1}{\sqrt{2}} (\sigma_1 + \sigma_3) = \frac{1}{\sqrt{2}}
\left(
\begin{array}{cc}
1& 1
\\
1 & -1
\end{array}
\right) = U^{-1}.
\label{unop}
\end{equation}
The Hilbert space can be written as a direct sum 
${\cal B}={\cal B}_+\oplus {\cal B}_-$ of the 
parity eigenspaces
of the parity operator $\hat{\Pi}=\sigma_1 \hat{\gamma}$ 
\cite{AMep,AMops,FG,Br}. 
Here $\hat{\gamma}=e^{i\pi \hat{a}^\dagger \hat{a}}$ 
induces {\em reflections} 
of the annihilation and creation operators:
$\hat{a}\rightarrow-\hat{a}$, 
$\hat{a}^\dagger\rightarrow-\hat{a}^\dagger$, 
and leaves the boson number operator 
$\hat{a}^\dagger \hat{a}$ invariant \cite{FG,Br}.
The corresponding 
parity eigenstates $\Phi^+$ and $\Phi^-$ of the eigenvalue
equation (\ref{eme}) 
contain one independent component each \cite{AMep,AMops,FG,Br},
\begin{equation}
\Phi^+(z) = \left(
\begin{array}{c}
\upvarphi^+
\\
\hat{\gamma}\upvarphi^+
\end{array}\right), \hspace*{0.8cm}
\Phi^-(z) = \left(
\begin{array}{c}
\upvarphi^-
\\
- \hat{\gamma}\upvarphi^-
\end{array}\right).
\label{pregs}
\end{equation} 
The respective parity eigenstates $\Phi^+(z)$ and $\Phi^-(z)$ 
satisfy the following eigenvalue equations for the 
independent (e.g. upper) component 
(cf. Eqs. (4.12-13) of Ref. \cite{FG})
\begin{eqnarray}
H^+\upvarphi^+ &=& [A + B\hat{\gamma} + C ]\upvarphi^+= E^+\upvarphi^+,
\nonumber
\\
H^-\upvarphi^- &=& [A - B\hat{\gamma} + C ]\upvarphi^-= E^-\upvarphi^-,
\nonumber
\end{eqnarray}
where $A=\omega_c \hat{a}^\dagger \hat{a}$, $B=\mu$, and
$C=g (\hat{a}^\dagger + \hat{a})$.
Here we have written $E^\pm$ since, in general, the spectra 
of $H^+$ and $H^-$ do not coincide.

Now, in the Bargmann space of entire functions, 
the action of $\hat{\gamma}$ becomes 
(Eq. (10) of Ref. \cite{AMops}; Eq. (37) of Ref. \cite{AMep})
\begin{equation}
\hat{\gamma}\upvarphi^\pm (z)=\upvarphi^\pm (-z) 
           =\sum_{n=0}^\infty (-1)^n \phi_n^\pm z^n.
\nonumber
\end{equation}
Thereby, the Rabi model can be characterized 
by a pair of the three-term recurrences 
(Eq. (37) of Ref. \cite{AMep})
\begin{eqnarray}
 \lefteqn{
 \phi_{n+1}^\pm +\frac{1}{\kappa (n+1)}\, 
              [n  - \upepsilon  \pm(-1)^n\Delta]\phi_{n}^\pm 
}\hspace*{3cm}
\nonumber\\
      &&    + \frac{1}{n+1}\, \phi_{n-1}^\pm = 0,
\label{rbmb2}
\end{eqnarray}
where $\upepsilon\equiv E^\pm/\omega_c$, 
$\Delta=\mu/\omega_c=\omega_0/(2\omega_c)$, and 
$\kappa=g/\omega_c$ reflects the coupling strength \cite{AMep}. 
Because the recurrence (\ref{rbmb2}) satisfies 
the conditions that guarantee uniqueness
of the minimal solution,
i.e. each $\upvarphi^\pm(z)$ generated by 
the respective minimal solutions 
is {\em unique}, the spectrum in each parity eigenspace 
${\cal B}_\pm$ is necessarily 
{\em nondegenerate} (cf. sec. 5.2 of Ref. \cite{AMops}).

As shown in our recent work \cite{AMops}, 
the substitution $\phi_n^\pm (\upepsilon)=P_{n}^{(-1)}(x)/n!$
transforms each of the two three-term recurrences (\ref{rbmb2})  
into the defining equation of
monic orthogonal polynomials [cf. Eq. (\ref{chi3trg})],
\begin{eqnarray}
P_n^{(\alpha)}(x) &=&(x-c_{n+\alpha})P_{n-1}^{(\alpha)}(x)
            - \lambda_{n+\alpha} P_{n-2}^{(\alpha)}(x),
\label{chi3tra}
\\
P_{-1}^{(\alpha)}(x) &=& 0,~~~~~~ P_{0}^{(\alpha)}(x)=1,
\nonumber 
\end{eqnarray}
where $\alpha=-1$, $x=\upepsilon/\kappa$, 
\begin{equation}
c_n \equiv \frac{1}{\kappa }\,[n \pm (-1)^n\Delta],
\label{cn}
\end{equation}
$\lambda_{n}=n$ for $n>0$, 
and $\lambda_{0}=1$ \cite{AMops}.
Note that the coefficients $c_n$ and $\lambda_{n}$ 
are {\em real} and independent
of $x$, and $\lambda_{n+\alpha} > 0$ for $n\ge 1$.
Because the Stieltjes measure $d\psi(x)$ 
in the orthogonality relations (\ref{bor})
has a {\em discrete} support \cite{AMops},
Eq. (\ref{bor}) reduces to 
\begin{eqnarray}
 {\cal L}[P_m(x)P_n(x)] & = & 
\sum_{x_i\in\Lambda} P_m^{(-1)}(x_i) P_n^{(-1)}(x_i)\, d\psi(x_i)
\nonumber\\
&& = n! \delta_{mn},
\label{rpor}
\end{eqnarray}
where $\Lambda$ is a one-dimensional lattice representing
the discrete support of $d\psi(x)$. 
Therefore, the resulting polynomials are {\em discrete} 
orthogonal polynomials \cite{NSU,KS,BFA,Clr}.
One can verify that, except for the special limiting case $\Delta=0$ (discussed below),
the polynomials are {\em non-classical} orthogonal polynomials 
(i.e. they cannot be recovered as solution of a second-order difference 
equation of {\em hypergeometric} type - cf. Secs. 2-3 of Ref. \cite{NSU};
the classical polynomials are called the Hahn class of orthogonal 
polynomials in Sec. V-3 of Ref. \cite{Chi}).

The three-term recurrences (\ref{rbmb2}) imply that the exponents
$\varsigma=0$, $\upsilon=-1$ 
and $\tau=\varsigma-\upsilon=1\geq 1/2$ [cf. Eq. (\ref{rcd})]. 
Therefore, the conditions required for the validity 
of our first to third observations are satisfied for the Rabi model,
provided that $\kappa< 1$. 
The above range encompasses not only the conventional {\em strong} coupling 
regime characterized in that $\kappa=g/\omega_c \lesssim 10^{-2}$ but also
the {\em ultrastrong} ($\kappa\gtrsim 0.1$) coupling regime, and 
overlaps with the {\em deep strong} ($\kappa\approx 1$) coupling regime \cite{CRL}.
For $\kappa \gtrsim 0.1$ the validity of 
the rotating wave approximation (RWA) breaks down and
the relevant physics can only be 
described by the full Rabi model \cite{Rb}.

\section{Main results}
\label{sc:mnr}
Let us elucidate our main results on the example
of the exactly solvable limit $\Delta=0$ describing
a {\em displaced harmonic oscillator} \cite{Schw}. 
The expansion coefficients $\phi_n$ are 
known to be determined by the associated Laguerre polynomials 
$L_n^{(\zeta-n)}(\kappa^2)$, 
where $\zeta=\upepsilon +\kappa^2=\kappa x +\kappa^2$.
(cf. Eq. (2.16) of Ref. \cite{Schw} and Sec. 4 of Ref. \cite{AMops}). 
Note in passing that energy variable $\zeta$ is not 
the polynomial variable of the associated Laguerre polynomials.
It is expedient to work with the (monic) Charlier polynomials \cite{Chr} 
(cf. Eqs. VI-1.4-5 of Ref. \cite{Chi}) and express $\phi_n$ as 
(see Sec. 4 of Ref. \cite{AMops})
\begin{equation}
\phi_n (\upepsilon) = \frac{P_n^{(-1)}(x;\Delta=0)}{n!} 
      = \frac{1}{n!\kappa^{n}}\, C_n^{(\kappa^2)}(\zeta).
\label{pnm1}
\end{equation}
The Stieltjes measure $d\psi^{(\kappa^2)}$ 
in the orthogonality relations of the Charlier polynomials 
(cf. Eq. VI-1.3 of Ref. \cite{Chi})
is known to be the step function
\begin{equation}
d\psi^{(\kappa^2)}(\zeta)
 =\sum_{l=0}^\infty \frac{e^{-\kappa^2} \kappa^{2\zeta}}{\zeta!}\, \delta(\zeta-l).
\label{chwf}
\end{equation}
Note in passing that $d\psi^{(\kappa^2)}$
is the {\em Poisson distribution function}
of probability theory at the jumps \cite{Chi}.
The jumps occur at $\zeta=0,1,2,\ldots$ The set of all the jumps forms 
the support of the Stieltjes measure $d\psi^{(\kappa^2)}$ \cite{Chi}, which in turn
is known to be formed by the set of all the limit zero points $\xi_l$ defined
earlier by Eq. (\ref{1p5p6}) \cite{Chi}. 
Thus the orthogonality relations are 
\begin{equation}
 \sum_{l=0}^\infty C_m^{(\kappa^2)}(l)C_n^{(\kappa^2)}(l)
\, d\psi^{(\kappa^2)}(l) = \kappa^{2n} n! \delta_{mn}.
\label{chpor}
\end{equation}
Not surprizingly, the location of jumps correspond exactly to 
the eigenvalues of the displaced harmonic oscillator \cite{Schw}
\begin{equation}
\upepsilon_l=l-\kappa^2
\label{bslnc}
\end{equation} 
(including $l=0$). The jumps define an {\em equidistant} lattice.
Because the orthogonality relation (\ref{chpor}) reduces to an infinite sum,
the Charlier polynomials are said to be classical {\em discrete} 
orthogonal polynomials on an equidistant lattice \cite{NSU,KS,BFA,Clr}.
The adjective {\em classical} implies that the Charlier polynomials 
can be recovered as solutions of a second-order difference 
equation of {\em hypergeometric} type (cf. Sec. 2 of Ref. \cite{NSU}).
\begin{figure}
\begin{center}
\includegraphics[width=\columnwidth,clip=0,angle=0]{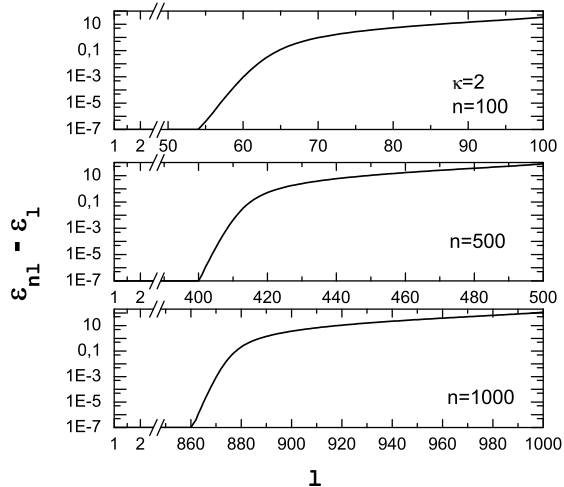}
\end{center}
\caption{
An illustration of the approximation of the spectrum 
in the case of the exactly 
solvable displaced harmonic oscillator, which corresponds to 
the Rabi model in the limit $\Delta=0$. Shown is the
difference of the approximants determined 
by the zeros $x_{nl}$, $l=1,2,\ldots,n$, 
of $\phi_n(\upepsilon)$ compared to
the exact eigenvalues $\upepsilon_{l-1}=l-1-\kappa^2$ 
for $\kappa=2$ and different degree $n$ of $\phi_n(\upepsilon)$. 
The precision in calculating zeros was 
set to seven decimal places.}
\label{fgzrsg2}
\end{figure}

Thus in the example of the displaced harmonic oscillator
our conclusions can be shown to be rigorously
valid also for $\kappa\ge 1$.
As a by-product, none of the zeros of $\phi_n$ coincides
with the exact spectrum.
In more detail, Eq. (74) of Ref. \cite{AMops} shows that $\phi_n$ is a sum of 
polynomials in the dimensionless energy parameter $\zeta$,
\begin{equation}
\phi_n = \sum_{j=0}^n (-1)^{n-j}
\frac{\kappa^{n-2j}}{(n-j)!j!}\, \prod_{k=0}^{j-1} (\zeta-k).
\label{cnsp}
\end{equation}
The spectral points $\zeta=l\in \mathbb{N}$ 
(including $l=0$) of the displaced harmonic oscillator  
are characterized by a sudden collapse of the
degree of $\phi_n$ to a polynomial of merely the 
degree $(l-1)$ in $\zeta$ for any $n\ge l$ \cite{AMops}.
Consequently, $\phi_n$ reduces for any $\zeta=l$ to a finite sum
of $l$ terms, each ranging from $(-1)^n \kappa^n/n!$ for $j=0$ 
to the $j=(l-1)$th term
\begin{equation}
(-1)^{n+1-l} \frac{l \kappa^{n+2-2l}}{(n+1-l)!}\cdot
\nonumber 
\end{equation}
Clearly, the points of the spectrum $\zeta=l$
{\em do not} coincide with the zeros of any of $\phi_n$.
However, each of the individual terms 
rapidly decreases with increasing $n$ 
in its absolute value down to zero. 
It is straightforward to show that for any $\zeta=l\in\mathbb{N}$
the absolute value of $\phi_n$ could be bounded by 
$l^2 \, \max(\kappa^n,\kappa^{n+2-2l})/(n+1-l)!$.
Hence
\begin{equation}
\phi_n(\zeta=l) \rightarrow 0 \hspace*{1.5cm}(n\rightarrow \infty),
\nonumber 
\end{equation}
i.e., at any given point of the spectrum $\phi_n$ rapidly vanishes
in the limit $n\rightarrow \infty$ down to zero
(cf. figure \ref{fgzrsg2}).

For a nonzero value of $\Delta$ our polynomials 
cannot be recovered as solution of a second-order difference 
equation of {\em hypergeometric} type
(cf. Secs. 2-3 of Ref. \cite{NSU}).
Thus $\Delta\ne 0$ induces a deformation of 
the Charlier polynomials to {\em non-classical} discrete
orthogonal polynomials and, at the same time, a deformation of the underlying 
equidistant lattice. Although neither the weight function
nor the deformed lattice are analytically known, 
the orthogonality relations (\ref{bor}) enable us to conclude that
the above deformation is a {\em norm preserving deformation}.
Indeed, Eq. (\ref{bor}) implies that the norm depends only on 
the value of the recurrence coefficients $\lambda_j$. However, the latter
do not depend on $\Delta$. Thus, as exemplified
by Eq. (\ref{rpor}), $||P_n^{(-1)}||^2=n!$ for any value of $\Delta$
[cf. Eqs. (\ref{pnm1}) and (\ref{chpor})].

\begin{figure}
\begin{center}
\includegraphics[width=\columnwidth,clip=0,angle=0]{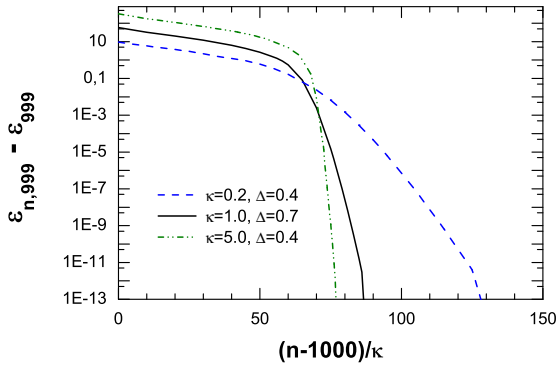}
\end{center}
\caption{Convergence of the $1000$th zero $\upepsilon_{n,999}$ of 
$\phi_n(\upepsilon)$ toward the exact $1000$th eigenvalue
$\upepsilon_{999}=998.907883759510,\, 997.950425260357$ 
and $973.989087026621$ of the Rabi model in the positive 
parity eigenspace for 
$(\kappa,\Delta)=(0.2,0.4),\, (1,0.7),\, (5,0.4)$, 
respectively. The ground state energy 
$\upepsilon_{0}$ corresponds to the first 
zero of $\phi_n(\upepsilon)$.
Each decreasing sequence $\upepsilon_{n,999}$ 
forms a discrete flow converging toward 
the corresponding exact eigenvalue down to machine precision.
}
\label{fgrb1000}
\end{figure}
Regarding the question {\bf Q1} raised in the preceding section,
the answer turns out thus not only be affirmative, but the result
exceeds all the expectations also in the case of the Rabi model 
(cf. figure \ref{fgrb1000}).
The convergence of the zeros to the spectrum is very fast.
The tail of highest-order zeros which do not
approximate the spectrum increases for a given $n$th level with 
increasing $\kappa$. 
Although we can rigorously prove our results only for $\kappa< 1$,
numerics strongly suggests that the main 
conclusions remain to be valid also for $\kappa\ge 1$.
Convergence of each discrete flow of zeros has been independently checked by the
Schweber quantization criterion \cite{Schw,AMep,AMops}
(see also Sec. \ref{sc:schw} below).
The example in figure \ref{fgrb1000} shows 
that the fraction $n_t/\kappa=(n-1000)/\kappa$ saturates at some 
constant value for $\kappa\gtrsim 1$.
The eigenstate corresponding to the $n$th eigenvalue can 
be approximated for $N=n+n_t$ as 
\begin{equation}
\upvarphi(z) \approx \sum_{l=0}^N \phi_l(x_{Nn}) z^l 
+ \sum_{l=N+1}^\infty \frac{(-\kappa z)^l}{l!}\cdot
\label{pss1}
\end{equation}
In agreement with the asymptotic behaviour of 
the {\em minimal} solution enforced by 
the Perron-Kreuser theorem (\ref{mins}),
$\phi_n(\upepsilon)$ has to behave for any eigenvalue as
$\phi_n(\upepsilon)\sim (-\kappa)^n/n!$ for sufficiently large $n$
(note that the three-term recurrences (\ref{rbmb2}) implies $a=1/\kappa$,
$\varsigma=0$, $b=1$, $\upsilon=-1$).

By the well known relations connecting the zeros and 
coefficients of a polynomial (Theorems I-4.2 and IV-3.1 of Ref. \cite{Chi})
\begin{equation}
\sum_{l=0}^{n-1} \upepsilon_{l} \approx \kappa \sum_{l=1}^n c_{l}
=\frac{n(n-1)}{2} + d,
\label{enrl}
\end{equation}
where we have substituted from (\ref{cn}) for $c_{l}$, and 
$d$ is one of $0,\pm \Delta$.
The latter justifies that energy eigenvalues are 
rather closely distributed around a straight line \cite{Ksc}.

The answer to our question {\bf Q2} appears peculiar.
Our computers allows us to work only in a {\em finite} precision.
However, in the given precision, the spectrum of the Rabi 
model can be determined by the zeros
of the polynomials $\phi_n(\upepsilon)$.
We have seen above that 
the discrete zeros flow has nowhere to flow than 
to the spectral point (cf. figure \ref{fgrb1000}).
Therefore, in any finite precision the Rabi model is
indistinguishable from an {\em algebraically} solvable model.
At the same time, only a computer with 
unlimited precision would recognize that the Rabi model is not
{\em algebraically} solvable, because the limit
$n\rightarrow\infty$ is required for the zeros
flow to converge to the spectrum. Note in passing (see below)
that the same limit is also required in Braak's solution.

\section{Discussion}
\label{sec:disc}
Solving the spectral condition (\ref{cd2}) implies
an entirely new, efficient, and relatively general method 
in determining the spectrum. The method 
differs both from (i) a brute force 
numerical diagonalization, (ii) searching for zeros of 
functions determined by infinite continued
fractions as in the Schweber method 
(cf. Eq. (A.16) of Ref. \cite{Schw}), and (iii) Braak's approach.
Only the lowest {\em 10-20} energy levels are within the reach 
of both Braak's solution \cite{Br3} and, as shown below, 
of the Schweber method \cite{Schw,AMep} - you are invited 
to convince yourself by running numerical F77 code that 
has been made available on-line \cite{AMr}. 
A brute force numerical diagonalization allows one 
to determine above {\em 2000} energy levels 
in double precision (cca 16 digits). However any deeper analytic
insight is missig. Note in passing that the presently calculable
{\em 1350} energy levels per parity subspace have been obtained
by the simplest stepping algorithm. Then the 
numerical limitation in calculating zeros 
are {\em over-} and {\em underflows}. 
Typically, with increasing $n$ the respective 
recurrences yield first increasing and then decreasing 
$\phi_n$. It is conceivable that the use of a more
sophisticated algorithm could overcome the limit
of the total number of calculable energy levels of ca. {\em 1350} levels
per parity subspace, or ca {\em 2700} levels for the 
Rabi model in total, in double precision.

That expansion coefficients $\phi_n$ could be determined
by orthogonal polynomials is, strictly speaking, not necessary
for working of our method based on solving the spectral 
condition (\ref{cd2}). The method could provide also
an efficient numerical way of obtaining the spectra of the models which expansion 
coefficients $\phi_n$ {\em cannot} be given
by orthogonal polynomials.

\subsection{Comparison with Braak's solution}
Braak \cite{Br} argued that a {\em regular} 
spectrum of the Rabi model in the 
respective parity eigenspaces
is given by the zeros of transcendental functions
\begin{equation}
G_\pm(\zeta)=\sum_{n=0}^\infty K_n(\zeta,\kappa)
\left[1\mp\frac{\Delta}{\zeta-n}\right]\kappa^n.
\label{sol}
\end{equation} 
The coefficients $K_n(\zeta,\kappa)$ are obtained recursively 
by solving the Poincar\'{e} difference equation 
\begin{equation}
K_{n+1} - \frac{f_n(\zeta)}{(n+1)}\,
             K_n + \frac{1}{n+1}\, K_{n-1}=0
\label{sa8}
\end{equation}
upwardly for $n\ge 1$, where
\begin{equation}
f_n(\zeta)=2\kappa+\frac{1}{2\kappa}
\left(n-\zeta - \frac{\Delta^2}{n-\zeta}\right),
\label{f-n}
\end{equation}
$\kappa$ and $\Delta$ are as in Eq. (\ref{rbmb2})
(cf. Eq. (A8) of Schweber \cite{Schw}, 
which has mistyped sign in front of his $b_{n-1}$, 
and Eqs. (4) and (5) of \cite{Br}).
The initial condition is
\begin{equation}
K_1/K_0=f_0(\zeta) = 2\kappa-\frac{1}{2\kappa}
\left(\zeta - \frac{\Delta^2}{\zeta}\right),
\nonumber 
\end{equation}
with $K_0$ being a normalization constant.
Braak's solution requires
(i) to solve for an undetermined number of complicated
functions $K_n(\zeta,\kappa)$ having poles
at discrete values of $\zeta$ (cf. Sec. 5.1 of Ref. \cite{AMops}),
(ii) to assemble the functions $K_n(\zeta,\kappa)$
into $G_\pm(\zeta)$ according to Eq. (\ref{sol}), 
(iii) to solve for zeros of $G_\pm(\zeta)$.
Thus it is not surprizing that Braak's approach
reaches its limits already 
at ca. {\em 20} energy levels in double precision \cite{Br3}.
Even if an additional analytic continuation step could increase
the number of calculable energy levels in Braak's approach 
to around {\em 100} \cite{Br3}, the number is still 
by an order of magnitude lower than what is possible 
within our approach.

In contrast, in our approach the structure of 
any $\phi_{n}(\upepsilon)$'s is clear - 
they are all determined by orthogonal polynomials. 
Further, only a {\em single} 
well-behaved $\phi_{N}(\upepsilon)$ of the degree $N=n+n_t$,
$n_t>0$, is required to determine any first $n$ eigenvalues 
of the Rabi model arranged 
in increasing order as zeros of $\phi_{N}(\upepsilon)$. 
Importantly, our approach also provides an efficient
registry of energy levels.
Indeed, a given $\phi_{n}(\upepsilon)$
has $n$ distinct real zeros. Therefore any omission 
of energy level can be easily identified.
The latter could be useful 
in any future statistical analysis of the spectra \cite{Ksc}. 
To reach unlimited precision, the limit
$n\rightarrow\infty$ is required. However, the latter 
is also necessary in the definition of $G_\pm(\zeta)$.

\subsection{Failure of Schweber's method for higher order eigenvalues}
\label{sc:schw}
It is believed that the spectrum of the Rabi model 
can be formally 
determined by the Schweber quantization criterion 
expressed in terms of infinite continued
fractions (cf. Eq. (A.16) of Ref. \cite{Schw} 
and Refs. \cite{AMep,AMops}),
\begin{equation}
0=F(x)\equiv a_0 + \frac{-b_{1}}{a_{1}-} \frac{b_{2}}{a_{2}-}
\frac{b_{3}}{a_{3}-}\cdots,
\label{fdfs}
\end{equation}
where [cf. Eq. (\ref{rbmb2})]
\begin{equation}
a_n =\frac{1}{\kappa (n+1)}\, 
              [n  - \upepsilon  \pm(-1)^n\Delta],~~~~~
b_n=\frac{1}{n+1},
\end{equation}
According to the Wallis formulas (Eqs. (III.2.1) of Ref. \cite{Chi};
Eqs. (4.2-3) of Ref. \cite{Gt}), the infinite continued fraction
in Eq. (\ref{fdfs}) can be expressed as the limit 
\begin{equation}
r_0=\lim_{n\rightarrow\infty} \frac{A_n}{B_n}\cdot
\label{r0l}
\end{equation}
Here $A_n$ and $B_n$ are the $n$th {\em partial numerator}
and the $n$th {\em partial denominator}, respectively.
We have shown that 
the ratio on the r.h.s. of Eq. (\ref{r0l}), 
also known as a {\em convergent},
can be expressed as the limit of 
the ratios of the polynomials \cite{AMops}
\begin{equation}
r_0=\lim_{n\rightarrow\infty} \frac{P_{n-1}^{(1)}(x)}{P_n(x)},
\label{r0lo}
\end{equation}
where $P_{n}^{(\alpha)}$ satisfy 
Eq. (\ref{chi3tra}) for $\alpha=0,1$.
The $n$th {\em partial numerator} $A_n$ in Eq. (\ref{r0l}) 
is related to $P_{n-1}^{(1)}(x)$, whereas the $n$th {\em partial denominator} 
$B_n$ is related to $P_{n}(x)$.
Any numerical method of 
computing $F(x)$ through Eqs. (\ref{r0l}) and (\ref{r0lo})
has to impose an unavoidable cutoff at some $n=N\gg 1$.
For any finite $n$ the ratio in (\ref{r0lo}) enables 
the partial fraction decomposition (PFD) 
(Theorem III-4.3 of Ref. \cite{Chi}),
\begin{equation}
 \frac{A_n}{B_n}=\frac{P_{n-1}^{(1)}(x)}{P_n(x)}=\sum_{l=1}^n
\frac{M_{nl}}{x-x_{nl}},
\label{pfdr}
\end{equation}
where the numbers $M_{nl}$ are all {\em positive}
and satisfy the condition $\sum_{l=1}^n M_{nl}=1$ \cite{AMops,Chi}.

Let $F_n(x)$ denote a finite-order approximation to $F(x)$
defined by Eq. (\ref{fdfs}), which is obtained by approximating 
$r_0$ in Eqs. (\ref{r0l}) and (\ref{r0lo}) by 
the PFD in Eq. (\ref{pfdr}).
One finds $dF_n(x)/dx <0$
whenever the derivative exists \cite{AMops}.
Consequently, $F_n(x)$ decreases from $+\infty$ to $-\infty$ 
between any two subsequent $x_{nl}<x_{n,l+1}$ and
there is exactly one zero of $F_n(x)$ \cite{AMops}
(cf. figs. 1,2 of Ref. \cite{AMep} 
and fig. 1 of Ref. \cite{AMops}). 
$F_n(x)$ has its zeros and poles {\em interlaced} 
on the real axis \cite{AMops}.
As is the case of any associated OPS's 
(see sec. III.4 of Ref. \cite{Chi}),
 the zeros of $P_{n}^{(\alpha)}(x)$ 
and $P_{n-1}^{(\alpha+1)}(x)$ are {\em interlaced} 
(Theorem III-4.1 of Ref. \cite{Chi}).
Specifically, 
\begin{equation}
x_{nl}^{(\alpha)}<x_{n-1,l}^{(\alpha+1)}<x_{n,l+1}^{(\alpha)},
\hspace*{0.8cm} \alpha=-1,0.
\label{3p4p1}
\end{equation}
(It is reminded that the superscript $\alpha=-1$ denotes the zeros
of $\phi_n$'s.)
The second of the rigorous sharp 
inequalities in Eq. (\ref{3p4p1}) implies
\begin{equation}
x_{n-1,l-1}^{(1)} < x_{nl} < x_{n+1,l+1}^{(-1)}.
\label{zridr}
\end{equation}
(For the sake of notation the superscript $(0)$
for $\alpha=0$ will be suppressed in what follows.)
The above scenario can be indeed confirmed 
numerically for a small number 
of the very first eigenvalues (cf. figs. 1,2 of Ref. \cite{AMep}; 
fig. 1 of Refs. \cite{AMops,AMcm}).

\begin{figure}
\begin{center}
\includegraphics[width=\columnwidth,clip=0,angle=0]{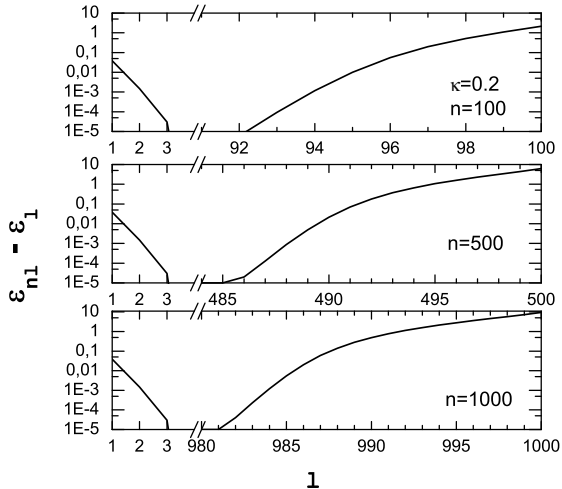}
\end{center}
\caption{
An approximation of the spectrum 
in the case of the exactly 
solvable displaced harmonic oscillator
by the zeros $x_{nl}$, $l=1,2,\ldots,n$, of the partial 
denominators $B_{n}$'s [cf. Eq. (\ref{r0l})].
Similar to figure \ref{fgzrsg2},
shown is the difference of the approximants 
$\upepsilon_{nl}$ 
compared to the exact {\em excited} eigenvalues 
$\upepsilon_{l}=l-\kappa^2$, $l>0$, 
for $\kappa=0.2$ and different degree $n$ of $B_n$. 
The ground state energy 
$\upepsilon_{0}$ at $l=0$ is not accounted for by the 
OPS for $\alpha=0$. The precision in calculating zeros 
was set to five decimal places.}
\label{fgzrs}
\end{figure}
Nevertheless, any practical implementation of the Schweber 
method {\em fails} for higher order eigenvalues. 
Depending on the model parameters, one can determine 
only up to {\em 10}-{\em 20} eigenvalues, and that already 
in the exactly solvable limit of the displaced 
harmonic oscillator - cf. F77 code made available online \cite{AMr}.
We have traced the failure down to a curious property 
of zeros of associated OPS - cf. data files \cite{AMdf}.
Surprisingly enough, after a first few of initial zeros
[e.g. beginning with $l\gtrsim 2$ for $(\kappa,\Delta)=(0.2,0.4)$]
one finds that, in spite of the strict inequalities (\ref{zridr}), 
\begin{equation}
 x_{n-1,l-1}^{(1)} \simeq x_{nl} \simeq x_{n+1,l+1}^{(-1)}.
\label{zrid}
\end{equation}
For $l\gtrsim 4$ and $(\kappa,\Delta)=(0.2,0.4)$ the zeros then
coincide up to more than {\em five} decimal places 
(provided that $n$ is sufficiently large) - cf. data files \cite{AMdf}.
Because of the coagulation of zeros (\ref{zrid}), any singularity 
of $F_n(x)$ becomes {\em numerically invisible}.
The coagulation can be undertood in that the position of
zeros for each the OPS's is largely determined by the respective sequences 
\begin{equation}
q^{(\alpha)}_n(x) 
= \frac{\lambda_{n+\alpha}}{(x-c_{n+\alpha})(x-c_{n+\alpha+1})}\cdot
\nonumber 
\end{equation}
However, in the present case one finds that
\begin{equation}
q^{(\alpha)}_{n-\alpha}(x) 
  =\frac{\kappa^2(n+1)}{(n-\upepsilon\pm (-1)^n\Delta)
         (n+1-\upepsilon\mp (-1)^n\Delta)} 
\nonumber 
\end{equation}
do not depend on the value of $\alpha$.
The coagulations of zeros is demonstrated in figure \ref{fgzrs}.
The latter shows that it is possible to determine
a large part of the spectrum by looking
at the zeros of the partial denominators $B_{n}$'s. 
Only marginally worse approximation property show
the zeros of the partial numerators $A_{n}$'s (not shown).
We emphasize that the failure of Schweber's method 
is entirely down to the finite precision 
of numerical calculations. Although the rigorous 
theory underlying Schweber's method \cite{Schw,AMep,AMops,AMcm}
is perfectly valid, a practical value of the method
may be thus rather limited. The above conclusions are expected 
to apply also to alternative continued fraction
expressions for the Rabi model studied by Ziegler \cite{Zg}.

\subsection{Algebraic solvability}
The Rabi model is a typical example of {\em quasi-exactly solvable} (QES) models 
in quantum mechanics \cite{TU,Trb,BD,KUW,KKT}. 
The QES models are distinguished by the fact 
that, for a chosen set of model parameters, a {\em finite} number of their eigenvalues and 
corresponding eigenfunctions can be determined {\em algebraically}
\cite{TU,Trb,BD,KUW,KKT}. In the case of the Rabi model \cite{Rb},
the latter eigenvalues correspond to the {\em Juddian} exact isolated
solutions \cite{Jd,Ks}. The possibility of any other polynomial
solution can be excluded by recent results of Zhang \cite{Zh}
(see also Sec. 4.2 of Ref. \cite{AMops}).
In agreement with the no-go theorem of Zhang \cite{Zh}, 
none of the zeros of any polynomial from any of 
the OPS $\{P_n^{(\alpha)}(x)\}$ coincides with the spectrum.
The spectral points could only be reached in 
the limit $n\rightarrow\infty$. 

Although the very notion of quantum integrability 
is the subject of ongoing dispute \cite{CM}, 
it is largely accepted that 
if eigenvalues can be determined {\em algebraically}
\cite{TU,Trb,BD,KUW}, this implies {\em integrability} 
and {\em solvability} \cite{CM}. Our results
indicate that the algebraic solvability
could be intricately linked with available precision.
Conceptually, and from a broader perspective, 
the above properties of the QES Rabi model provide an example of 
that, numerically, there may be only very subtle 
difference between {\em exactly} and {\em quasi-exactly} solvable models, 
if the latter are characterized by discrete orthogonal polynomials.
In general, any (i.e. not necessary QES) model that satisfies 
the spectral condition (\ref{cd2}) and is characterized 
by discrete orthogonal polynomials could exhibit such 
a solvability.

\subsection{Relation to the Jaynes and Cummings model and the effect of the RWA}
For dimensionless coupling strength $\kappa=g/\omega_c \lesssim 10^{-2}$,
the physics of the Rabi model is known to be well captured by the
analytically solvable Jaynes and Cummings (JC) model \cite{JC}. 
The latter is obtained from the former upon applying 
the rotating wave approximation (RWA), whereby the 
coupling term 
$\sigma_1 (\hat{a}^\dagger + \hat{a})$ in Eq. (\ref{rabih}) 
is replaced by $(\sigma_+ \hat{a} + \sigma_-\hat{a}^\dagger)$, 
where $\sigma_\pm \equiv (\sigma_1 \pm i \sigma_2)/2$.
The eigenstates of the JC model are linear combinations of 
the product states $|\psi_{1n}\rangle=|n\rangle|e\rangle$ and 
$|\psi_{2n}\rangle=|n+1\rangle|g\rangle$ in the Hilbert space
${\cal B}=L^2(\mathbb{R})\otimes\mathbb{C}^2$,
where the respective $|e\rangle=(1,0)^t$ and $|g\rangle=(0,1)^t$, 
with the superscript $t$ indicating the transpose,
stand for the excited and ground state
in the spin space $\mathbb{C}^2$ \cite{Schw,JC}.
The product states $|\psi_{1n}\rangle$ and $|\psi_{2n}\rangle$ form 
a basis of an invariant subspace of the operator
\begin{equation}
\hat{J}=\mathds{1} \hat{a}^\dagger \hat{a}
         +\frac{1}{2}\,(\mathds{1}+\sigma_3),
\nonumber 
\end{equation}
which generates a continuous $U(1)$ symmetry of the JC model \cite{Br,JC}.
The invariant subspace is characterized by the eigenvalue $n+1$ of $\hat{J}$.
In terms of the parity operator $\hat{\Pi}=-\exp(i\pi \hat{J})$,
each invariant subspace of $\hat{J}$ is positive or negative parity
subspace depending on if $n$ is even or odd, respectively. 
Therefore, in each invariant subspace of $\hat{J}$, and hence 
for the eigenstates of the JC model,
the parameter $w=\pm (-1)^n\Delta$ in Eq. (\ref{rbmb2}) reduces to $w=\Delta$.

If $\kappa$ becomes small, the solution of the JC model \cite{JC} suggests to arrive 
at approximate solutions of Eq. (\ref{rbmb2}) by setting all but two subsequent
expansions coefficients $\phi_l$ and $\phi_{l+1}$ to zero.
By forming corresponding $\varphi_\pm(z)=\phi_lz^l+\phi_{l+1}z^{l+1}$ according 
to Eq. (\ref{pss}), substituting into Eq. (\ref{pregs}), and unitary transforming
by the operator $U$ given by Eq. (\ref{unop}), one can verify that
the corresponding parity eigenstates $\Phi^+$ and $\Phi^-$ [cf. Eq. (\ref{pregs})]
in the single-mode boson picture $\hat{H}_{sb}$ become 
\begin{equation}
\Phi(z) = \left(
\begin{array}{c}
\phi_l z^l
\\
\phi_{l+1} z^{l+1}
\end{array}\right)
\nonumber 
\end{equation} 
in the conventional representation $\hat{H}_R$. Here we have used that
$\Phi$ derives from the positive or negative parity
eigenstate depending on if $l$ is even or odd, respectively.
Note in passing that $\Phi(z)$ is yet undetermined linear combination 
of the JC states $|\psi_{1l}\rangle$ and $|\psi_{2l}\rangle$ 
in a given invariant subspace of $\hat{J}$.
Obviously, upon imposing RWA onto $\hat{H}_R$ in Eq. (\ref{rabih}) and 
substituting our $\Phi(z)$ as trial wave functions one would recover
the JC model solution.

In order to investigate the effect of the RWA on the exact solution,
we determine the eigenvalues of the JC model from the exact equations.
Upon considering Eq. (\ref{rbmb2}) for $n=l$ and $n=l+1$ one arrives at
\begin{eqnarray}
\frac{\phi_{l+1}}{\phi_{l}} &=& - \frac{1}{\kappa(l+1)}\, [l-\upepsilon +\Delta],
\nonumber\\
\frac{\phi_{l}}{\phi_{l+1}} &=& - \frac{1}{\kappa}\, [l+1-\upepsilon -\Delta].
\label{jceq}
\end{eqnarray}
One can recast Eqs. (\ref{jceq}) in the matrix form
\begin{equation}
\left(
\begin{array}{cc}
 l+\Delta   & (l+1)\kappa 
\\
 \kappa  &  l+1-\Delta
\end{array}
\right)\left(
\begin{array}{c}
\phi_{l}
\\
\phi_{l+1}
\end{array}
\right)
=\upepsilon \left(
\begin{array}{c}
\phi_{l}
\\
\phi_{l+1}
\end{array}
\right).
\nonumber 
\end{equation}
The secular equation reduces to a quadratic equation
\begin{equation}
\upepsilon^2 -(2l+1)\upepsilon +l(l+1) + \Delta - \Delta^2 - \kappa^2 (l+1) = 0.
\nonumber
\end{equation}
The eigenvalues are 
\begin{equation}
\upepsilon_{\pm}=l+\frac{1}{2} \pm \frac{1}{2} \sqrt{1-4\Delta +4\Delta^2 + 4\kappa^2 (l+1)}.
\nonumber
\end{equation}
Given $\Delta=\omega_0/(2\omega_c)$, one finds
\begin{equation}
1-4\Delta +4\Delta^2 = 1- \frac{2\omega_0}{\omega_c} + \frac{\omega_0^2}{\omega_c^2} 
=\frac{(\omega_0- \omega_c)^2}{\omega_c^2}=\delta_c^2,
\nonumber
\end{equation}
i.e. square of the normalized detuning parameter 
$\delta_c=(\omega_0 - \omega_c)/\omega_c$ of the JC model \cite{JC}.
Note in passing that $\delta_c\ll 1$ in the RWA, because the latter
is reliable only if $g|\omega_0-\omega_c|\ll \omega_0,\, \omega_c$.
The eigenvalues can be thus recast as
\begin{equation}
\upepsilon_{\pm}=l+\frac{1}{2} \pm \frac{1}{2} \sqrt{\delta_c^2 + 4\kappa^2 (l+1)},
\label{jcel}
\end{equation}
which is the familiar form of the eigenvalues of the JC model \cite{JC}.

Any exact regular solution of the Rabi model is 
characterized by {\em infinite} set of {\em nonzero} expansion coefficients
$\phi_n$, which for sufficiently large $n$ behave as
$\phi_n \sim (-\kappa)^n/n!$ [cf. the Perron-Kreuser theorem (\ref{mins}) and 
the recurrence Eq. (\ref{rbmb2})].
Interestingly, the RWA takes implicitly into account the effect of 
$\phi_n\ne 0$ for $n\ne l, l+1$.
If the coefficients were ignored, Eq. (\ref{rbmb2}) for $n=l-1$ and $n=l+2$ would 
require that additionally 
\begin{equation}
\phi_{l} = 0, \hspace*{1.2cm} \phi_{l+1}/(l+3) =0.
\nonumber
\end{equation}
Afterwards one would find for $\upepsilon=\upepsilon_+$
\begin{equation}
\frac{\phi_{l+1}}{\phi_{l}} = \tan \frac{\theta}{2} = \frac{2\kappa}{D +\delta_c}
       =\frac{D -\delta_c}{2(l+1)\kappa},
\end{equation}
where we have substituted from (\ref{jcel}) for $\upepsilon$, denoted 
$D=\sqrt{\delta_c^2 + 4\kappa^2 (l+1)}$, and used that $\Delta-(1/2)=\delta_c/2$.
Because
\begin{equation}
\tan \theta = \frac{2 \tan \frac{\theta}{2}} {1 - \tan^2 \frac{\theta}{2}},
\nonumber %
\end{equation}
and
\begin{equation}
\tan^2\frac{\theta}{2}=\frac{D - \delta_c}{D +\delta_c}\frac{1}{(l+1)},
\nonumber %
\end{equation}
one can determine $\tan \theta$ as 
\begin{equation}
\tan \theta = \frac{4\kappa}{D + \delta_c} \frac{(l+1)[D + \delta_c]}{lD +(l+2)\delta_c}
= \frac{4(l+1)\kappa}{lD +(l+2)\delta_c}\cdot
\nonumber %
\end{equation}
The latter would not coincide with $\tan \theta =2\kappa\sqrt{l+1}/\delta_c$ 
for the JC model solution \cite{JC}. 
Hence if one tries from the very outset to ignore 
in the exact solution 
all but a pair of expansion coefficients, one will arrive at 
the RWA energies but not to the RWA tangent value.

\subsection{Open problems}
In Sec. \ref{sc:rbm} it has been alluded to that our 
polynomials underlying the Rabi model are {\em non-classical} {\em discrete} 
orthogonal polynomials \cite{NSU,KS}. Various generalizations 
of the classical discrete orthogonal polynomials have been studied 
in the literature. However, they have been almost exclusively concerned with 
various generalizations of the Stieltjes weight function
while maintaining an underlying lattice on which the polynomials
are defined \cite{BFA,Clr}. An ensuing problem has been to determine 
the recurrence coefficients \cite{BFA,Clr}.

In the case of the Rabi model, the recurrence coefficients are 
explicitly known [cf. Eqs. (\ref{chi3tra}), (\ref{cn})]. 
As we have seen in Sec. \ref{sc:mnr}, the lattice is equidistant 
only in the special limiting case $\Delta=0$. Then the polynomials 
are proportional to the Charlier polynomials, which are related to each other 
according to (cf. Eq. (VI-1.7) of Ref. \cite{Chi})
\begin{equation}
\Delta_+ C_n^{(\kappa^2)}(x) = n C_{n-1}^{(\kappa^2)} (x),
\label{chrc}
\end{equation}
where $\Delta_+ u(x) = u(x+1)-u(x)$ denotes the forward finite 
difference operator \cite{NSU}. Their weight function (\ref{chwf}) 
satisfies a special form of the {\em Pearson} difference equation
\begin{equation}
\Delta_+ u(x) = \frac{\kappa^2 -x-1}{x+1}\, u(x).
\label{preq}
\end{equation}
A non-zero value of $\Delta$ induces a {\em norm-preserving} deformation of 
the Charlier polynomials to {\em non-classical} discrete orthogonal 
polynomials and, at the same time, a nonuniform deformation 
of the underlying equidistant lattice. 
The spectrum of the Rabi model is then nothing but 
the discrete nonuniform lattice.
The so-called $q$-analogs of the Charlier polynomials on {\em nonuniform} lattices
with the lattice points $x(n)=\exp(2wn)$ and $x(n)=\sinh(2wn)$, respectively,
with $w>0$ being some parameter, have been discussed in Sec. 3.6 of Ref. \cite{NSU}. 
Yet those extensions are still classical orthogonal polynomials that
are distinguished by the fact that all their properties are unambiguously determined
by the second-order difference equation of {\em hypergeometric} type 
which they satisfy \cite{NSU}.
Contrary to the main line of research of the discrete orthogonal 
polynomials community \cite{BFA,Clr}, the problems here 
are (i) to characterize the class of {\em non-classical} {\em norm-preserving} 
extensions of the classical discrete Charlier polynomials which encompasses 
the polynomials presented here (e.g. in terms of a suitable second order 
difference equation), (ii) to find generalizations
of Eqs. (\ref{chrc}) and (\ref{preq}) for $\Delta\ne 0$,
and (iii) to determine the nonuniform lattice 
on which the polynomials are defined. The problems are typically 
intertwined, because the forward finite difference operator is expected to 
operate on $\Lambda$ (cf. Sec. 3 of Ref. \cite{NSU}), 
whereas a second order difference equation arises on combining 
a relation of the type (\ref{chrc}) with the 
defining three-term recurrence (cf. Sec. VI-1 of Ref. \cite{Chi}).
The required extension has to be such that the average density of lattice points
is substantially preserved 
[note that Eq. (\ref{enrl}) only marginally depends on $\Delta$].

\section{Conclusions}
\label{sec:conc}
The spectrum of the Rabi model was shown to
coincide with the support of the discrete Stieltjes 
integral measure in the orthogonality 
relations of recently introduced non-classical discrete 
orthogonal polynomials.
This finding brings about a novel method 
of solving the Rabi, and similar to it, models. 
In the case of the Rabi model the method resulted in 
an analytic solution that is considerably simpler 
than Braak's solution \cite{Br,Br3}.
The eigenfunctions can be determined in terms 
of orthogonal polynomials, whereas the
eigenvalues are found as the polynomial zeros.
Thus any omission of an energy level
could easily be identified.
The simplicity of our analytic solution was
rewarded by the fact that the number of ca. {\em 1350} 
calculable energy levels per parity subspace 
in double precision obtained by a simple stepping
algorithm is almost two orders of magnitude higher 
than is possible to obtain by means of 
Braak's solution \cite{Br,Br3}.
A valuable insight as to whether a model 
is integrable or chaotic is provided by 
the energy level statistics.
Our results suggest that energy eigenvalues are 
rather closely distributed around a straight line \cite{Ksc}. 

Although we can rigorously prove 
our results only for $\kappa< 1$, numerics 
and exactly solvable example suggest that the main conclusions are
valid also for $\kappa\ge 1$.
Our results could thus provide a reliable point of departure
for the calculation 
of the dynamics of the Rabi model and its long-time evolution 
for all values of the dimensionless coupling $\kappa$.
The latter could be important to a great 
variety of physical systems, 
including cavity and circuit quantum electrodynamics, quantum
dots, polaronic physics and trapped ions \cite{KGK,BGA,FLM,NDH}. 
With new experiments rapidly approaching the limit of
the {\em deep strong} coupling regime $\kappa \gtrsim 1$, 
one expect such systems to open up a rich vein of research on truly 
quantum effects with implications for quantum 
information science and fundamental quantum optics \cite{CRL,KGK}.

\section{Acknowledgment}
Continuous support of MAKM is largely acknowledged.



\begin{thebibliography}{99}


\bibitem{Schw}
S. Schweber, 
Ann. Phys. (N.Y.) {\bf 41}, 205 (1967).


\bibitem{Brg}
V. Bargmann, 
Comm. Pure Appl. Math. {\bf 14}, 187 (1961).


\bibitem{AMep}
A. Moroz, 
Europhys. Lett. {\bf 100}, 60010 (2012).


\bibitem{Zh1}
Y.-Z. Zhang,
arXiv:1304.3979 [quant-ph].


\bibitem{AMops}
A. Moroz, to appear in Ann. Phys. (N.Y.) 
(arXiv:1302.2565).


\bibitem{Rb}
I. I. Rabi, 
Phys. Rev. {\bf 49}, 324 (1936).


\bibitem{Gt}
W. Gautschi, 
SIAM Review {\bf 9}, 24 (1967).


\bibitem{Chi}
T.~S. Chihara, 
An Introduction to Orthogonal Polynomials
(Gordon and Breach, New York, 1978).


\bibitem{NSU}
A. F. Nikiforov, S. K. Suslov, V. B. Uvarov, 
Classical Orthogonal Polynomials of a Discrete
Variable (Springer, Berlin, 1991).


\bibitem{KS}
W. Koepf and D. Schmersau,
J. Comput. Appl. Math. {\bf 90}, 57 (1998).


\bibitem{TU}
A. V. Turbiner and A. G. Ushveridze,
Phys. Lett. A {\bf 126}, 181 (1987).


\bibitem{Trb}
A. V. Turbiner,
Commun. Math. Phys. {\bf 118}, 467 (1988).


\bibitem{BD}
C. M. Bender and G. V. Dunne,
J. Math. Phys. {\bf 37}, 6 (1996).


\bibitem{KUW}
A. Krajewska, A. Ushveridze, and Z. Walczak,
hep-th/9601088.


\bibitem{KKT}
R. Koc, M. Koca, and H. T\"{u}t\"{u}nc\"{u}ler,
J. Phys. A: Math. Gen. {\bf 35}, 9425 (2002).


\bibitem{FG}
R. L. Fulton and M. Gouterman, 
J. Chem. Phys. {\bf 35}, 1059 (1961).


\bibitem{Br}
D. Braak,
Phys. Rev. Lett. {\bf 107}, 100401 (2011).


\bibitem{BFA}
L. Boelen, G. Filipuk, and W. Van Assche,
J. Phys. A: Math. Theor. {\bf 44}, 035202 (2011).


\bibitem{Clr}
P. A. Clarkson, 
J. Phys. A: Math. Theor. {\bf 46}, 185205 (2013).


\bibitem{CRL}
J. Casanova, G. Romero, I. Lizuain, 
J. J. Garc\'{\i}a-Ripoll, and E. Solano, 
Phys. Rev. Lett. {\bf 105}, 263603 (2010).


\bibitem{Chr}
C. V. L. Charlier,
Ark. Mat. Astr. Fys. {\bf 2}, 1 (1905-6).


\bibitem{Ksc}
M. Kus,
Phys. Rev. Lett. {\bf 54}, 1343 (1985).


\bibitem{Br3}
D. Braak,
Ann. Phys. (Leipzig) {\bf 525}, L23 (2013).


\bibitem{AMr}
The source code can be freely downloaded from 
\verb|http://www.wave-scattering.com/rabi.html|.


\bibitem{AMcm}
A. Moroz, 
arXiv:1205.3139 [quant-ph]. 


\bibitem{Zg}
K. Ziegler,
J. Phys. A: Math. Theor. {\bf 45},  452001 (2012).


\bibitem{AMdf}
Data files *gp2dp4n300.dat 
obtained for $(\kappa,\Delta)=(0.2,0.4)$ and $n=300$
are available from 
\verb|http://www.wave-scattering.com/rabi.html|.


\bibitem{Jd}
B. R. Judd,
J. Phys. C: Solid State Phys. {\bf 12}, 1685 (1979).


\bibitem{Ks}
M. Kus,
J. Math. Phys. {\bf 26}, 2792 (1985).


\bibitem{Zh}
Y.-Z. Zhang,
J. Phys. A: Math. Theor. {\bf 45}, 065206 (2012).


\bibitem{CM}
J.-S. Caux and J. Mossel,
J. Stat. Mech. P02023 (2011).


\bibitem{JC}
E. T. Jaynes and F. W. Cummings,
Proc. IEEE {\bf 51}, 89 (1963).


\bibitem{KGK}
G. Khitrova, H. M. Gibbs, M. Kira, S. W. Koch, and A. Scherer,
Nat. Phys. {\bf 2}, 81 (2006).


\bibitem{BGA}
J. Bourassa, J. M. Gambetta, A. A. Abdumalikov, Jr., 
O. Astafiev, Y. Nakamura, and A. Blais,
Phys. Rev. A {\bf 80}, 032109 (2009).


\bibitem{FLM}
P. Forn-D\'{\i}az, J. Lisenfeld, D. Marcos, 
J. J. Garc\'{\i}a-Ripoll, E. Solano, 
C. J. P. M. Harmans, and J. E. Mooij,
Phys. Rev. Lett. {\bf 105}, 237001 (2010).


\bibitem{NDH}
T. Niemczyk, et al, 
Nat. Phys. {\bf 6}, 772 (2010).



\end{thebibliography}
\end{document}